\documentclass{article}
\usepackage{graphicx}
\usepackage{fullpage}
\usepackage{subfigure}

\begin{document}

\title{Numeric Experiments in Relativistic Thermodynamics:\\
 A Moving System Appears Cooler}

\author{Constantin Rasinariu \thanks{E-mail: crasinariu@colum.edu}\\
Department of Science and Mathematics\\
Columbia College Chicago, Chicago, USA}


\maketitle

\begin{abstract}

In this paper we simulate a two dimensional relativistic ideal gas by implementing a relativistic elastic binary collision algorithm. We show that the relativistic gas faithfully obeys J\"uttner's speed distribution function. Furthermore, using this numeric simulation in conjunction with the energy equipartition theorem for a relativistic gas, we conclude that a moving system appears cooler.
\end{abstract}

\section{Introduction}

There is a long standing controversy on the rendering of the fundamental thermodynamic concepts in the special-relativistic form.

In 1907 Mosengeil, Planck, and Einstein  \cite{Mosengeil:1907,Planck:1907,Einstein:1907}, independently showed that if a thermodynamic system with temperature $T$ is at rest in the inertial frame $K$, then its temperature $T^{\,\prime}$ measured by an inertial frame  $K^{\,\prime}$ moving with speed $u$ relative to $K$, is $T^{\,\prime} = T / \gamma$ where $\gamma = (1-u^2)^{-1/2}$ (in $c=1$ system of units). Therefore, a moving system appears cooler. However, a completely opposite result, $T^{\,\prime} = T\,\gamma$, was proposed later by Blanu\u{s}a \cite{Blanusa:1947}, Einstein (in a little known letter to von Laue) \cite{Liu:1992}, Ott \cite{Ott:1963}, Arzelies \cite{Arzelies:1969} and many others. Then, Einstein \cite{Liu:1992} (in another letter to von Laue), Landsberg \cite{Landsberg:1966}, and others, pondered whether the temperature should be a relativistic invariant. Few years later, Balescu \cite{Balescu:1968} explored the possibility that all relativistic transformations are in fact equivalent via some sort of ``gauge'' transformations. This list is manifestly incomplete. The relativistic thermodynamics, in its special-relativistic formulation, has a long and convoluted history. A detailed discussion about this curious episode of modern physics is beyond the scope of this paper. The interested reader should consult for example \cite{Arzelies:1969, Guessous:1970, Liu:1992, Landsberg:2004} and the references therein.

It was often argued that because of the high temperatures, and the relativistic speeds involved, there are no experimental results that could settle the controversy on the relativistic thermodynamics. However, by deploying a realistic computer model, we can actually ``experimentally'' decide  on this question. Using a numeric model of a two dimensional relativistic ideal gas, in this paper we show that
\emph{a)} the relativistic gas faithfully obeys J\"uttner's speed distribution \cite{Juttner:1911}, and
\emph{b)} defining the temperature via the relativistic equipartition theorem, we show that the numeric simulations strongly favor the Mosengeil, Planck, and Einstein relativistic formula. Thus, we have a compelling numerical indication that moving objects should appear cooler.

\section{Computer Simulations}
We consider a two-dimensional ideal relativistic gas whose particles experience elastic collisions. The gas is enclosed inside a rectangular container, with perfectly reflecting walls, which conserve the energy-momentum of the colliding particles. Throughout this paper we will use a system of units where the speed of light, the mass of the particles, and the Boltzmann's constant are all taken equal to one $c=1, m = 1, k_B=1$.

The thermodynamic system is at rest in the lab when the container is at rest, and the average velocity of its particles is zero
$
\langle \vec{v}\ \rangle = 0
$.
If the gas is in thermal equilibrium at temperature $T = 1/\beta$, its speed distribution is given by J\"uttner's formula  \cite{Juttner:1911, Synge:1957}
\begin{equation}
\label{eq:Juttner}
f(v) = v \gamma_v^{4} \exp{\left( -\beta \,  \gamma_v \right)}\,/Z~,
\end{equation}
where $\gamma_v = (1-v^2)^{-1/2}$ is the Lorentz factor, and $Z =  e^{-\beta}(1+\beta)/\beta^2 $ is a normalization constant, chosen such that $\int_0^1 f(v) dv = 1$. Then, the average energy is given by
$
\langle E \rangle \equiv \int_0^1 E f(v) dv = (\beta^2 + 2\beta + 2)/(\beta^2+ \beta)
$
.
We will use this information to find the reciprocal of the temperature of the system at rest, from measuring its average energy. We get
\begin{equation}
\label{eq:temp}
\beta = \left(2-\langle E \rangle  + \sqrt{\langle E \rangle  ^2  + 4\,\langle E \rangle  -4 }\,\right) /\, (2\langle E \rangle  -2)~.
\end{equation}

\subsection{Numeric results}
For the experimental setup we considered $100,000$ identical particles, which initially,  were randomly distributed inside a rectangular box, and given the same speed in arbitrary directions. During one time step the algorithm checks for possible collisions with \emph{a)} the wall of the container (the box has the edges parallel with $x$ and $y$ axes), and \emph{b)} with another particle.
If none occurs, the particle advances one time step with the same velocity.

\medskip
\noindent a) {\em Collisions particle--wall:} If the particle hits the left or right wall, it reflects around the $x$-axis. If it hits the top or the bottom wall, the particle reflects around the $y$-axis.

\begin{figure}[htb]
\centering
  \includegraphics[width=130pt]{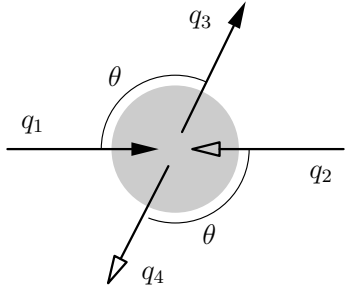}
  \caption[120pt]{The interaction as seen in the center of mass of the two particles. Here $q_1$ and $q_3$ are the momenta of the incoming and respectively outgoing particle $1$. Similarly, $q_2$ and $q_4$ are the momenta of the incoming and respectively outgoing particle $2$. The angle $\theta$ between incoming and outgoing particles is chosen randomly.}
  \label{fig:inter}
\end{figure}

\medskip
\noindent b) {\em Collisions particle--particle:} Point-like particles have practically no chance to collide if only contact, zero-range interactions are present. To bypass this difficulty, we define an ``interaction area'' around each particle as follows. If $(x_i, y_i)$ are the coordinates of particle $i$ as measured by an observer at rest with respect to the container, and if particle $j$ arrives at $(x_j, y_j)$ such that $(x_i-x_j)^2 +  (y_i - y_j)^2 \le R^2$, then the two particles ``collide''.  The distance $R$ is chosen to be much smaller than the linear dimensions of the container. The algorithm ``locks'' the two particles when colliding, thus ignoring the possibility of a third particle to participate in collision. This simplifies the simulation and does not introduce errors greater than a one percent. Knowing the ``in'' states of the particles entering the collision we determine the ``out'' states by using only the conservation of the energy-momentum. However, in this step, which is solved in the center of mass of the two particles, one has to choose an ``interaction angle'' $\theta$ for the outgoing particles, as illustrated in figure (\ref{fig:inter}).
Since particles are modeled as zero-size points experiencing only elastic collisions, $\theta$ is not meaningfully defined. Therefore, we use a pseudo-random number generator to assign a random value to the ``interaction'' angle $\theta$ .

\begin{figure}[htb]
\centering
  \includegraphics[width=250pt]{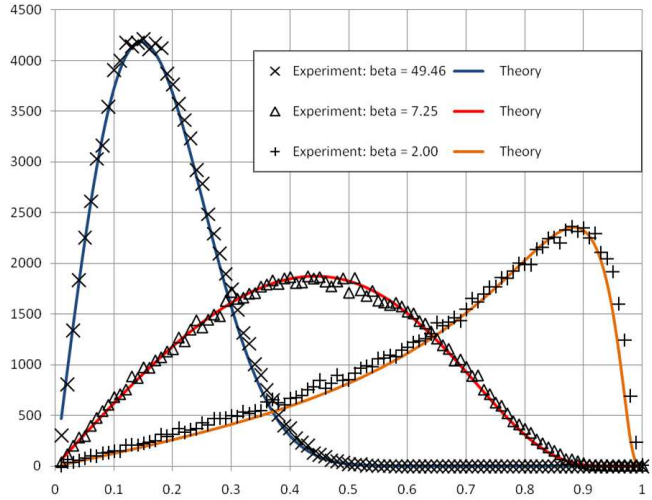}
  \caption{J\"uttner's speed distribution curves vs. the experimental speed distribution for a relativistic 2-d gas at reciprocal temperatures $\beta=49.46$, $\beta=7.25$, and $\beta=2.00$ respectively. The area under each distribution curve is equal to the total number of particles, $N = 100,000$.}
  \label{fig:exp}
\end{figure}

We monitored the average energy per particle, and considered that the system reached  thermal equilibrium when the average energy per particle fluctuated by less than $0.001\%$. From the onset of the experiment, the equilibrium was usually achieved after $40,000$ time steps. After the system reached thermal equilibrium, we determined its temperature by means of eq.(\ref{eq:temp}). By plugging the measured value back into eq.(\ref{eq:Juttner}) we were able to compare the theoretical curves with the histograms of the speed distribution obtained from  our computer modeling.  In figure (\ref{fig:exp}) we illustrate the experimental versus the theoretical results for the 2-d relativistic gas. The figure collects data from three different experiments: low temperature (with the corresponding reciprocal temperature $\beta\equiv mc^2/k_BT = 49.46$), intermediate temperature
($\beta=7.25$), and respectively high temperature (with the corresponding reciprocal temperature $\beta=2.00$). The agreement with J\"uttner's distribution is remarkable in all three cases.

\subsection{The relativistic energy equipartition}
Using the J\"uttner distribution function (\ref{eq:Juttner}) we can prove that for the system at rest, we obtain \cite{Guessous:1970, Landsberg:1992}
\begin{equation}
\label{eq:part}
1/\beta \equiv T = \langle p_x\, v_x \rangle = \langle p_y\,v_y \rangle~.
\end{equation}
We checked experimentally the validity of formula (\ref{eq:part}) for all  three reciprocal temperatures $\beta = 49.46~,7.25$ and $2.00$.  The results are summarized in table 1,
\begin{table}[ht]
\centering
\begin{tabular}{c c c}
$\beta~~~$      & $\beta_x~~~$  & $\beta_y~~~$  \\[0.5ex]
\hline
2.0000  & 2.0120    & 2.0147    \\
7.2476  & 7.27541   & 7.2487    \\
49.4561 & 49.6334   & 49.3251   \\[0.5ex]
\hline
\end{tabular}
\caption{Energy equipartition for the relativistic gas}
\label{table:part}
\end{table}
where we have denoted $\beta_x = \langle p_x\, v_x \rangle $ and $\beta_y = \langle p_y\,v_y \rangle$. Note that the numeric simulations show a good agreement with the theory.

\subsection{The relativistic transformation of temperature}
The system is in translation with the speed $\vec{u}$ with respect to the lab if the average velocity of all its particles measured by a stationary observer in the lab is $\langle \vec{v}\, \rangle = \vec{u}$.
Without loss of generality we will consider the system moving along the $x$-axis. Extending the equipartition theorem to the moving system $K^{\,\prime}$, we obtain
\begin{equation}
\label{eq:part1}
1/\beta^{\,\prime} \equiv T^{\,\prime} = \langle p_x^{\,\prime} \left(v_x^{\,\prime} - u \right) \rangle
= \langle p_y^{\,\prime}\,v_y^{\,\prime} \rangle~.
\end{equation}
For a complete proof of eq. (\ref{eq:part1}) the reader should consult the reference \cite{Guessous:1970}. Equation (\ref{eq:part1}) is a good tool to decide which relativistic transformation temperature formula is consistent with this experiment. As noted in \cite{Yuen:1970}, because the thermodynamic system is an extended object, the output of the measurements depends whether the measurement is performed at constant time in the rest frame or at constant time in the moving system. The extended geometrical dimensions of the thermodynamic system implies that a constant time in the rest frame corresponds to a set of times $t'$ in the moving system and vice-versa. We measured the particles in the container at constant time $t$ with respect to its rest frame.

Planck's formalism gives $\beta_{Planck}^{\,\prime} = \beta \,\gamma_u$, while Ott's formalism gives $\beta_{Ott}^{\,\prime} = \beta / \gamma_u$, where $\gamma_u = (1-u^2)^{-1/2}$ is the usual Lorentz factor.

\begin{figure}[htb]
  \centering
  \subfigure[The relativistic transformation of temperature for a relative low rest reciprocal temperature $\beta=49.46$.
  ]{\label{fig:cold}
  \includegraphics[width=200pt]{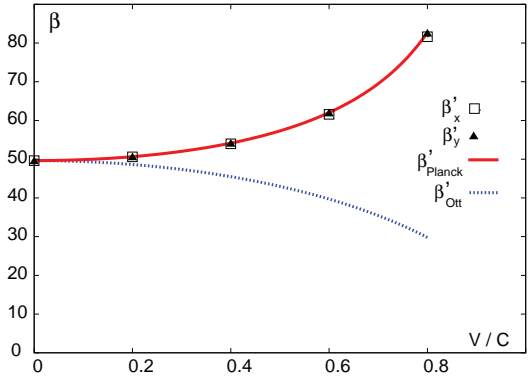}}
\hspace{.2in}
  \subfigure[The relativistic transformation of the temperature for  relative medium ($\beta=7.25$) and high ($\beta=2.00$) rest reciprocal temperatures.]{
  \label{fig:hot}
  \includegraphics[width=200pt]{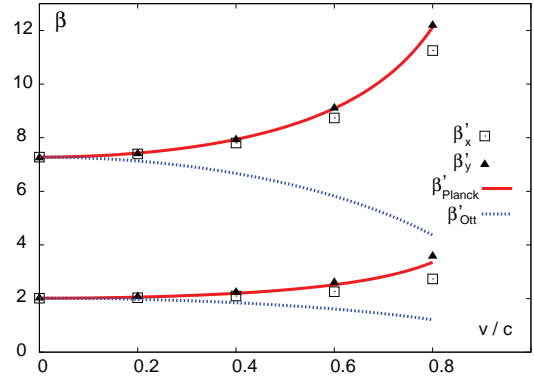}}
  \caption{Note that for the boost speeds $u = 0.2,~ 0.4,~ 0.6$, and $0.8$ the agreement between $\beta_x^{\,\prime}, \beta_y^{\,\prime}$ and $\beta_{Planck}^{\,\prime}$ is remarkable. We observe that with the increase of the temperature, $\beta_x^{\,\prime}$ shifts slightly below the values of $\beta_y^{\,\prime}$. In contrast, $\beta_{Ott}^{\,\prime}$ is clearly diverging from the experimental trend.}
\end{figure}

Choosing the same rest reciprocal temperatures as before ($\beta = 49.46,~7.25$ and $2.00$), we tested eq. (\ref{eq:part1}) for the boost speeds $u = 0.2,~ 0.4,~ 0.6$, and respectively $0.8$. The results are presented in figures (\ref{fig:cold}) and (\ref{fig:hot}). The robustness of the numeric algorithm was checked by repeating the experiments with different form factors for the rectangular box.

\section{Summary}
In this paper we solved the non-trivial problem of relativistic two-dimensional molecular dynamics. We showed that a relativistic ideal gas faithfully satisfies J\"uttner's speed distribution function. Using as guidance the relativistic equipartition theorem, we showed experimentally that the numeric simulations strongly favor the Mosengeil, Planck, and Einstein relativistic formula. Thus, moving objects appear cooler.

\section*{Acknowledgements}
We thank Asim Gangopadhyaya, Ovidiu Lipan and Mircea Pigli for valuable comments and discussions.

\end{document}